\documentclass{sistedes}

\usepackage{graphicx} 
\usepackage{cite}

\begin{document}

\bibliographystyle{plain}

\title{Digging into CTM's consciousness:\\
A possible mechanism for CTM generating self-conscious }

\author{
Shaoyang Cui
\and
Shanglin Wu
\and
Nikolai Madlener
}

\institute{
Peking University, No.5 Yiheyuan Road Haidian District, Beijing, P.R.China\\
}

\maketitle

\small

\begin{abstract}
    Based on the former work Conscious Turing Machine, in this paper, we attempt to talk about the consciousness of CTM, dig deeper into the self-consciousness in CTM, offer a clear definition of it, and design a possible model of the Model-of-the-World processor. To prove the consciousness of CTM does exist, we chose two definitions of human consciousness and extracted four key points to see if the CTM framework meets with them. If it does, we affirm that it's more likely to be able to generate consciousness. About self-consciousness, our definition of it refers to both the definition of conscious awareness in CTM and former studies about the duality of self. After that, we give a brief introduction to a possible model of MoTW processors including five important parts: Modeling function, Gist function, Value function, Cache, and Long term memory. Finally, we use some illusions and disorders to explain our MotW processor model, trying to understand how these illusions work on a CTM.

\end{abstract}

\section{Introduction}
The desire for a better understanding of consciousness comes out not only from philosophers and psychologists but also from computer scientists. For example, in the paper, \emph{A theory of consciousness from a theoretical computer science perspective: Insights from the Conscious Turing Machine}\cite{CTMPaper}, authors propose a formal TCS(theoretical computer science) model called Conscious Turing Machine.\\
As an important part of consciousness, self-consciousness receives high attention and has gradually become a hot spot in the field of artificial intelligence. With the common sense of human being as the only intelligent life with “self-consciousness”, only human's self-consciousness can be taken as a model in order to build AI with self-consciousness. This shows the meaning of better understanding self-consciousness from a computer science perspective.
In this paper, we are going to discover what is self-consciousness like in the CTM model. But before we start our trip, we need to review basic terminologies and concepts in CTM.




\subsection{Conscious Turing Machine(CTM)}
Conscious Turing Machine(CTM) is a conscious model inspired by the Global Workspace Theory of Bernard Baars\cite{Globalworkspace}.
Before the introduction of CTM construction, we prefer to introduce the definition of chunk first for the reason that it plays a key role in CTM's life.\\
A chunk is a 6-tuple, 
$<$ address, t, gist, weight, intensity, mood $>$,\\ where \textbf{address} is the producer processors' index, \textbf{t} stands for time, \textbf{gist} carries the information of this chunk. \textbf{Weight},\textbf{intensity} and \textbf{mood} are numbers used for competition. In a nutshell, a chunk is a small piece of information with a weight given by processors.\\
The CTM is a device that is defined by a 7-tuple,\\
$<$ STM, LTM, Up-Tree, Down-Tree, Links, Input, Output $>$, where
\begin{itemize}
    \item [1)]\textbf{STM} stands for Short Term Memory designed according to the Global Space theory which can only store one chunk.
    \item [2)]\textbf{LTM} stands for Long Term Memory, it's a set of processors which keep producing chunks and sent them to the competition tree (Up-tree).
    \item [3)]\textbf{Up-Tree} is a competition tree where chunks compete to enter the STM. For a probabilistic CTM, chunks with higher weight have a higher probability to enter the STM.
    \item[4)]\textbf{Down-Tree} is more like a broadcaster who sends the winner chunk of Up-tree to all LTM processors.
    \item[5)]\textbf{Link} is a special way for processors to communicate with each other. They could convey chunks through it instead of relying on the broadcast procedure.
    \item[6)]\textbf{Input} is what the CTM get from the real world while \textbf{Output} is the CTM's output to the real world.

\end{itemize}

While a CTM is working, all those LTM processors would be busy producing chunks and sending them to the Up-tree. Chunks compete to enter the STM and got broadcast to all processors. Some linked processors would convey chunks to each other as well. Processors adjust their algorithm according to received chunks and keep producing until the finite lifetime of CTM end. The interesting thing is that based on Blums' definition and design, a CTM like that could generate consciousness just like human beings.

\section{Conscious in CTM}
Since that we attempt to find the essence of the self-consciousness of a CTM, it's necessary to present a clear and correct definition of conscious in CTM. In this section, firstly we are going to briefly introduce the definition of CTM's consciousness given by Blums. After that, we'll introduce some former recognized studies and theories of human consciousness.\\
If the consciousness of CTM is well-designed, then it must be compatible with at least most of the existing definitions of human(or animal) consciousness. So we mapped those conscious theories into CTM's consciousness to see if they could work properly (or to say, those human conscious theories can be appropriately explained under CTM's structure). In that case, we bear out the rationalization of CTM's consciousness.

\subsection{Definition}
 Noticed that what we care about is the self-consciousness of the Conscious Turing Machine so its definition might sound inappropriate for human brains. And before presenting our definition, we prefer to offer some crucial details about the Conscious Turing Machine. \\
The CTM is a simplified yet powerful model. Theoretically, a completely and properly implemented CTM is sure to generate consciousness. Blums have given some interesting definitions of consciousness\cite{CTMPaper}, which fit the construction of CTM fair enough.\\

\begin{itemize}
\item[$\bullet$] \textbf{Conscious Content:} a chunk that wins the competition tree and reaches STM.\\
\item[$\bullet$] \textbf{Conscious Awareness:} the winner chunk broadcast to all LTM processors and gets received. \\
\item[$\bullet$] \textbf{Conscious Stream:} the sequence of the reception of winner chunks of LTM processors.\\
\end{itemize}
 It is important that only when LTM processors received the chunks can we say that consciousness is generated because, without the processors' action and response to the chunk, the broadcast itself is just an unconscious transmission. \\

 From this definition, we can figure out that although the LTM processors are unconscious, they are essential for generating consciousness. Those processors are just like neurons in the human brain, each of them is unconscious, but without them, there wouldn't be any chance to get consciousness. What's more, conscious awareness is the result of the procedure that STM and LTM work together instead of the specific activity of some particular processors. So through this vital procedure, we step over the boundary between unconscious and conscious.\\

\subsection{Proof of Conscious in CTM}
All those mentioned above are just definitions without theoretical supporting. They might sound right, but what we are pursuing is a more credible theoretical basis of consciousness. So based on some former studies about consciousness, here we present our proof of the rationalization of those definitions.\\

Our main approach is to check existing research and works that define "conscious" and extract some necessary functions and requirements for generating consciousness. We will then explore whether the framework of CTM can meet these requirements. If it does, we consider it a model capable of generating consciousness, and also, demonstrating the correctness of our definition of the conscious of CTM. We believe that this approach is correct to proof to the rationalization of consciousness within the CTM framework.\\

However, it's really hard to find a clear definition of consciousness. A great many famous philosophers and psychologists have put forward so many definitions and explanations, but there isn't a widely accepted one till now. Thomas Negal\cite{Bat} made a good analogy by imagining the experience of being a bat to aid in the comprehension of the concept of consciousness, but did not provide a clear definition."\\
The reason why those definitions are misty is that we can hardly test what consciousness is directly through experiments. Nonetheless, there are still some reasonable standards to define the conscious: starting with \textbf{duality}\\ 

\subsubsection{Dichotomously understand consciousness}
\ 
\newline
\indent Back in ancient Greek, Aristotle emphasize the distinction between awake and knocked out\cite{Aristotle}, which makes the concept of conscious and unconscious similar to perceiving and not-perceiving. Seems that the discussion about consciousness turns into a kind of cognitive problem. And if we go further through Aristotle's perspective, here is a more convincing definition of consciousness presented by Max Velmans\cite{UnderstandingConsciousness}\\
"A person, or other entity, is conscious if they \textbf{experience} something."\\
Velmans offered a constitution of the very pivotal concept experience as follows: 
\begin{itemize}
    \item[$\bullet$] the experience (of an entity) commonly associated with ourselves: thoughts, feelings, dreams...
    \item[$\bullet$] the experienced 3-dimensional world
\end{itemize}
According to the definition, we come up with that a conscious agent should be able to \textbf{model and interact with the real world} and also have the ability of \textbf{ knowing itself}, can \textbf{experience special activities like thoughts, feelings, and dreams}. We can figure out that the realization of all those functions mentioned above is the integration of many simple functions.  \\
Just similar to Velmans' theory, Stanislas Dehance presents his special comprehension in his paper(What is consciousness, and could Machines have it?),which divided the consciousness into two separated parts:\\
\begin{itemize}
    \item[1)] \textbf{Global Availability of Relevant Information.}
    \item[2)] \textbf{Self-Monitoring.}
\end{itemize}
According to Dehance's definition of consciousness, a conscious entity should be able to \textbf{handle information both globally and locally}(depends on which way is better) and could \textbf{monitor itself} based on information from both the environment and the entity itself.\\
The question is: Could a CTM realize all those requirements for generating consciousness? Just in terms of the current CTM construction, the answer is undoubted yes. We'll respectively discuss these items in detail.\\
\begin{itemize}
    \item \textbf{The ability of perception and comprehension.}\\
    As a CTM is equipped with enough sensors and processors, it's capable to get a perception of the real world: via visual, auditory, tactile, and other kinds of input. With the help of a powerful LTM processor called Model-of-the-World, it can understand the world and even establish an approximated simplified simulated world in its "brain". That's the same ability for a CTM to make a dream. When it comes to 'thoughts' and 'feelings', we tend to believe that they could be generated by processors and are contained in the most significant part of a chunk: the gist. A further discussion about this will include an in-depth consideration of \textbf{the Branish}, which is not what we're focused on here in this paper. At least, we're pretty sure that a CTM can perceive and comprehend. \\ 
    \item \textbf{The ability to be aware of itself.}\\
    Also, as designed, a CTM has enough fine-grained sensors and processors to get and understand plenty of information from itself. And all those processors would be trained to produce chunks according to a common value function that aims to help the CTM 'live' better through its lifetime, in other words, a CTM will always attempt to do things that are beneficial to itself. Based on those sensors and processors, it's not hard for a CTM to distinguish itself from other entities via modeling itself in its \textbf{outer world}.  \\
    \item \textbf{The ability to handle data both globally and locally.}\\
    A CTM can surely deal with data both globally and locally. By \textbf{globally} we mean the CTM invokes all processors to handle the data(or to say, the data is available to all processors) and in contrast, by \textbf{locally} we mean it only invokes some of the processors. There isn't an explicit boundary between 'globally' and 'locally' in CTM. All those data come in the form of chunks conveyed to LTM processors at time $t$. After all, processors have produced new chunks, those chunks will be sent to the competition Up-tree to battle for entering the STM. It's clear that in a probabilistic CTM, those chunks' weights are comparable in the competition without any unit conversion, therefor we believe all processors are using a common system of units to calculate the weight of each chunk. That indicates the CTM model can globally handle information. The ability of processing information locally is hidden in the whole processing-competition procedure. For example, if a CTM only wants to handle chunks from specific processors, it would just give those irrelevant chunks from other processors much lower weight so that those processors almost have no chance to 'hand in' their chunks to the STM, which is equivalent to make those irrelevant processors 'sleep'(stop producing chunks) for a while.
    \item \textbf{The ability of self-monitoring}\\
    \[Predictive\_Dynamics = Prediction + Feedback + Learning\]
    To prove that a CTM can self-monitoring, we should discuss the Sleeping Expert Algorithm(SEA). It seems that the SEA is specially designed for this constraint. As the formula is given above, all the LTM processors will continuously carry on the SEA, predicting the value of each possible assignment of the chunk, getting feedback from received chunks, and adjusting its algorithm to diminish and correct errors. This prediction-feedback-learning cycle starts at CTM's birth and last for its whole lifetime aiming at helping CTM 'live' better in the 'world'. That's just what we want for self-monitoring.\\
\end{itemize}

All those LTM processors are unconscious, and, of course, the STM itself is unconscious too. So neither the STM nor the LTM could generate "experience"(conscious) on its own. \\ In more detail, the generation of experience includes both sensors receiving inputs and processors handling input data. Those processors need to "communicate" with each other(via STM's broadcast and links) to process information correctly. That's why we define conscious awareness as a sequence of specific procedures instead of a single activity. 

Some researchers attempt to detect the origin location of consciousness in our brain via neuroscience measures. But till now we couldn't find a specific region(or some groups of regions) of the brain that is in charge of generating consciousness (like the visual cortex). This fact verifies our point that the generation of consciousness is a process rather than some specific neural activity of some particular processors from another perspective. \\

As George Miller once said: "Consciousness is a world-worn smooth by millions of tongues." We can hardly map all existing conscious theories into CTM's construction. 
However, considering Velmans' and Dehance's definition of conscious, based on the discussion above, we can come up with the conclusion that the CTM framework can meet all four extracted requirements. That's why we believe CTM is a powerful conscious model with great universality which could generate consciousness for certain and the definition of its conscious is rational.

\section{Definition of self-conscious in CTM}
Since that we've discussed some former conscious theories and proved that CTM stands a good chance of having consciousness based on its design. We want to find out whether a CTM could generate self-consciousness, furthermore, propose its possible mechanism.\ 
Just like consciousness, we've got tons of definitions of self-conscious, like 'conscious of conscious'\cite{CouldRob}. Exactly none of them involved its underlying mechanism. And unfortunately, the fact is that our 
species' pride seems to harm our study of self-consciousness. When we meet someone else, we can easily tell that he or she is self-conscious, but when it comes to animals like cats or dogs, there is a controversy. Since ancient times, philosophers tend to associate self-consciousness with spirit, which is a uniquely human property. \\

\subsection{Self-conscious awareness is a procedure}
\ 
\newline
However, among various self-conscious theories there is one thing in common: Almost none of them split 'conscious' and 'self-conscious' apart and discuss them respectively, which indicates that although there is no conclusive evidence, people think self-consciousness has a close association with consciousness.\\
The relationship between consciousness and self-consciousness is hard to tell. We couldn't just affirm that self-consciousness is a superior kind of consciousness or consciousness is a simplified form of self-consciousness. However, we firmly believe that they are not two irrelevant concepts because self-consciousness is sure to have all those properties of consciousness we've discussed in \textbf{3.2}. \\
What's more, from the point of the initial formation of consciousness, it's ridiculous that two concepts similar like this were generated in totally different ways. That may overcomplicate the brain and cause unwanted waste of computation and energy.\\
However, our ultimate goal is to seek the secret of the generation of consciousness and self-consciousness in CTM but not simulate the actual brain activities. If an explanation is rational, we can just assume it's correct.\\
So in our definition, both conscious and self-conscious are generated by the same procedure.\\

\subsection{The duality of Self-conscious}
\
\newline
The duality of self might be a good point to start with. Williams James\cite{ThePofP} thought that 'myself' should be divided into two parts: 'me' and 'I' where 'me' stands for an objective self and 'I' stands for a subjective self. In more detail, Max Velmans\cite{UnderstandingConsciousness} presented that our self-experiences should be divided into 'body experiences' and 'Inner experiences'. The 'Body experience' consists of interoceptive sensation, kinaesthesis, bodily pleasure and pain, and so on, while the 'inner experience' should involve thoughts, memories, feelings... \\
Speaking of subjective self, we can assume that thoughts would be produced as the outputs of LTM processors and be stored in every relative chunk's gist in Branish and memories would be stored in the processors' memory. So that processors indeed could be the key difference between self-conscious awareness and conscious awareness. \\
Self-consciousness requires an objective part: body experience. That means the processor should be capable to get information about CTM itself. In addition, generating self-consciousness is not an ordinary task. It requires processors to be aware of the CTM's inner state(including both body experience and inner experience mentioned above) together with the information of the outer environment, furthermore, it ought to have its value orientation to generate thoughts properly according to the CTM's current state for the entity's further development.\\

\subsection{\textbf{MIT}}
Before offering a clear definition, let's compare self-conscious awareness to conscious awareness. Since we've accepted the fact that they are generated by the same procedure, there must be some details that are different.\\ 
We believe that the distinction of processors in these two similar procedures plays a vital role in the generation of self-consciousness. As it's known to us all, each processor is equipped with its function so that it could carry on some different jobs and tasks. So that our goal is to find one(or a group of) proper processors that can satisfy the requirements of being self-conscious.\\
However, there isn't any single processor equipped with all those abilities. So we propose a powerful group of special processors: the \textbf{MIT}, where, 
\begin{itemize}
    \item \textbf{M} is the Model-of-the-world processor(\textbf{MoTW}).
    \item \textbf{I} is the Instruction-Generator processor(\textbf{IG})
    \item \textbf{T} is the Thought-Generator processor(\textbf{TG})
\end{itemize}
Based on \textbf{MIT}, we propose a simple definition of self-consciousness awareness in the CTM. 

\subsection{Definition of CTM's self-conscious}
\ 
\newline
As we argued, the formation of self-conscious awareness in CTM should be a procedure instead of a single activity of any processor. Since self-conscious and conscious are two analogical concepts, we assume that they are generated by the same procedure. Also, the duality of self-consciousness requires the CTM could be aware of both the subjective self and objective self, which refers to the ability to be aware of the state itself and generate applicable gists(instructions, thoughts...). And we designed a group of special processors--the \textbf{MIT} to meet those constraints.\\
Based on the definition of conscious content and conscious awareness, we here present the definition of \textbf{Self-conscious Content} and \textbf{Self-conscious Awareness} as follows:\\
\begin{itemize}
    \item[$\bullet$] \textbf{Self-conscious Content:} a chunk that wins the competition tree and reaches STM, in addition, this chunk should be made by \textbf{MIT}(But not all chunks made by MIT could be self-conscious content).\\
    \item[$\bullet$] \textbf{Self-conscious Awareness:} the Self-conscious Content broadcast to all LTM processors and gets received. \\
\end{itemize}
With this definition, we are going to dig deeper into the mechanism of CTM's self-consciousness: a possible model of MIT.

\section{MotW, IG, and TG}
As we defined above, some special MIT chunk would be the 'hero' of self-conscious awareness. So it's meaningful to find which kinds of component is essential for MIT to fulfill its functions.\\
Since that there is no concrete implementation of CTM, we'll just start with the human brain, assuming that the human brain is a neuron-based CTM to find how the MoTW works.  

\begin{figure}[htp]
    \centering
    \includegraphics[width=10cm]{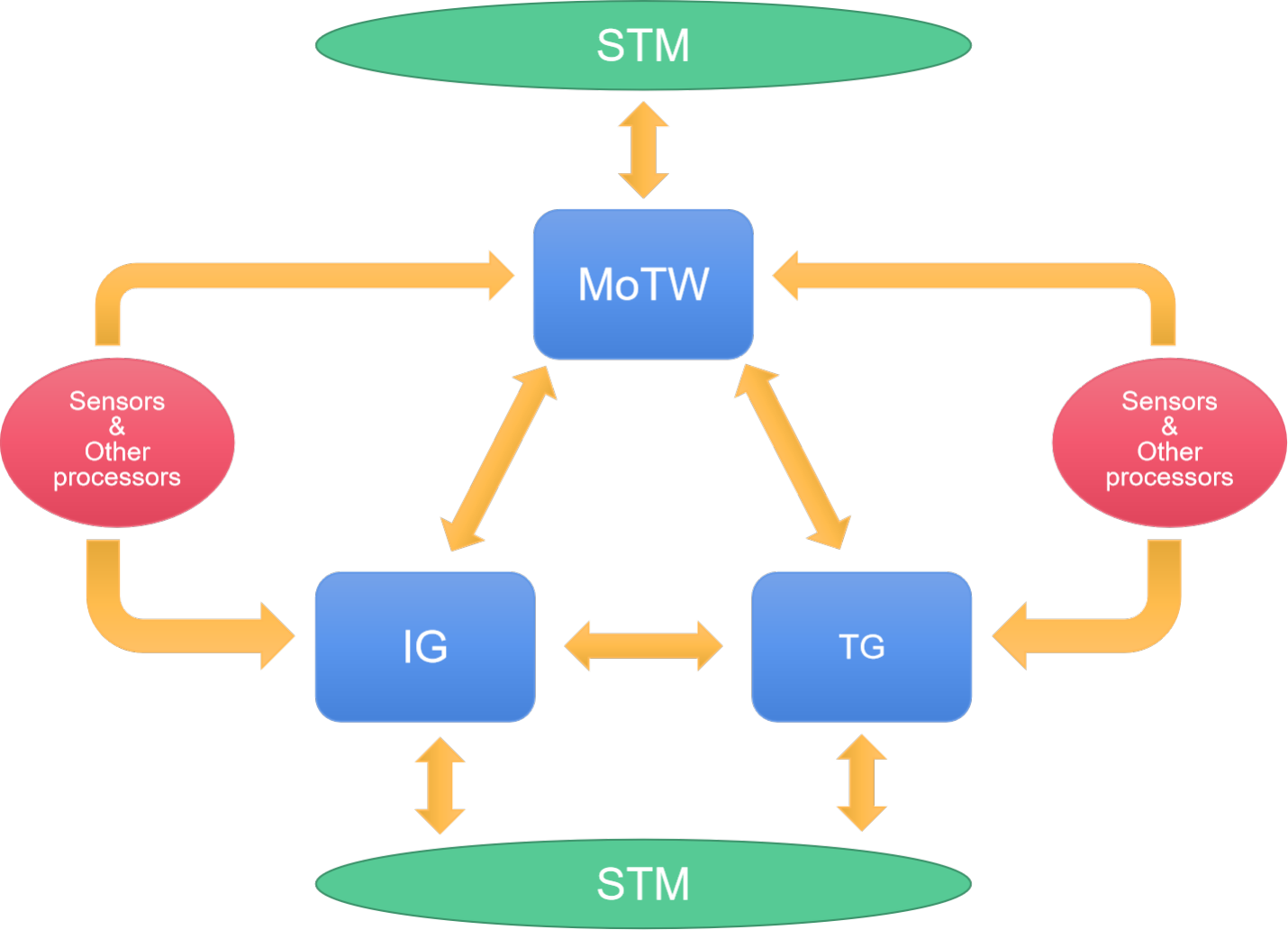}
    \caption{a) The yellow arrows in the figure represent the transmission of chunks between processors and STM. b) We have omitted the communication between other processors and STM in this figure. }
    \label{fig:misunderstanding}
\end{figure}

Here we present a schematic diagram of the MIT architecture, where the most crucial components depicted are three processors(the Model-of-the-World processor, the Instruction Generator processor, and the Thoughts Generator processor) and the links between them. Within the MIT group, we believe that the links between the three processors are of utmost importance due to the need for close information exchange among them. Broadcasting via STM is inefficient for this purpose.\\

We want to clarify here that although the CTM paper defines that MoTW can directly send output maps to the actuators, we believe that this function should be separated from MoTW and a new processor should be defined to implement it. MoTW already can model the world, which is powerful enough. Integrating two such powerful functions into one processor is not practical. In addition, we believe that there should be some delay between understanding the world and taking action, which is likely composed of the time for information transmission between different processors. If these two behaviors are integrated into one processor, it would erase this delay and treat the entity's perception of the environment and response as inputs and outputs of a multivariate function, with a delay of at most the function's runtime. That is unreasonable.\\

 In the forthcoming paragraphs, we will propose several rational functions of MIT to figure out its mechanism. \\

\subsection{MotW: Modeling the world}

The MoTW processor can receive information about itself and the environment from both sensors and other processors to correctly construct a model of the real world. \\
The most exciting function of the MoTW processor is generating the model\textbf{inner world} and \textbf{outer world}, where\\
\begin{itemize}
    \item \textbf{inner world} stands for the model of CTM itself, while
    \item \textbf{outer world} stands for the model of the real world.
\end{itemize}
MoTW is capable to be aware of the state of both itself and the environment. That's one of the most important functions helps CTM establish its thorough comprehension of the real world. Next, we will propose an M function that could show how MotW models the world.\\

\subsubsection{Modeling function \textbf{M} }
\ 
\newline
One of the MoTW's main functions is to model both itself and the environment. Hence we designed a Modeling function \textbf{M}:
\[[inner\_world, outer\_world] = M(S, K) = a_s*S + a_k*K\]
\begin{itemize}
    \item \textbf{S} stands for sensation, it comes from all sensors of CTM including the sense of visual, audio, tactile..., noticed that all those in a CTM were conveyed via binary numbers. The \textbf{M} function would transfer them into other modalities.
    \item \textbf{K} stands for the CTM's knowledgement(we made up this word to describe a concept, which is a combination of memory and knowledge) of the world. It includes both a CTM's knowledge and memory of the world. Noticed that the knowledgement could be MoTW's own as well as knowledgement of other processors received from broadcast or link. This parameter is gradually acquired by the CTM instead of being equipped at birth.
    \item \textbf{$a_s$, $a_k$} stands for the weight of sensation and knowledgement.
\end{itemize}
We'd like to offer more explanations of our definition here. It's easy to understand that the modeling function has an indispensable demand for sensations of all modalities to model the world. But the role that knowledgement plays in modeling is equally essential. \\
Dichotomously talking about knowledgement, we divide it into knowledge and memory. Of course, that memory is a significant part. For instance, even if we are not at home(which means we couldn't get any sensations of home directly), we can still model all those rooms according to our memory. However, the effect of knowledgement is not so explicit. We prefer to explain it with an example of visual illusion. 
\begin{figure}[htp]
    \centering
    \includegraphics[width=6cm]{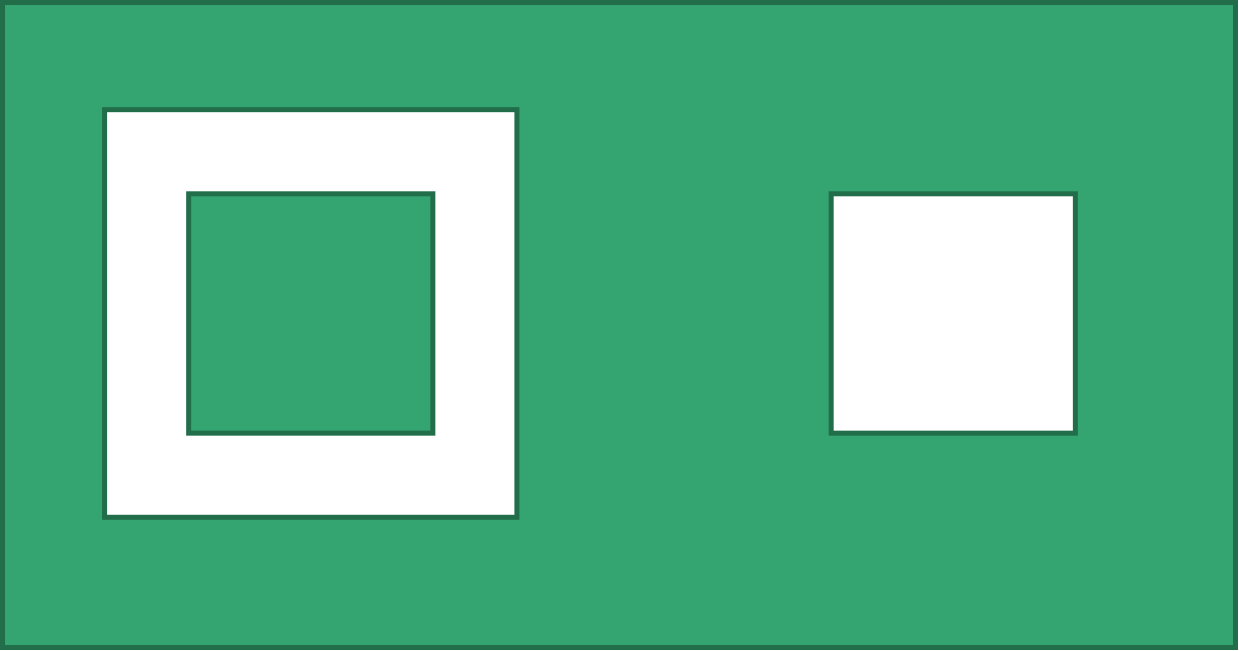}
    \caption{An image of a visual illusion}
    \label{fig:misunderstanding}
\end{figure}
In Fig.1., there are two squares in different colors, and most people's intuition is that the green one on the left side is bigger than the white one. But the fact is they are equivalent. \\
Our point is that the existing knowledgement of the world would correct the modeling function's wrong intuition. Furthermore, after being told that they are the same size, these two squares would seem to be the same size. That's how knowledgement helps model the world.\\
However, sometimes even though we know there is a visual illusion, what we see won't change our knowledgement. For instance, the famous Zöllner illusion shown in Fig.2.\\

\begin{figure}[htp]
    \centering
    \includegraphics[width=6cm]{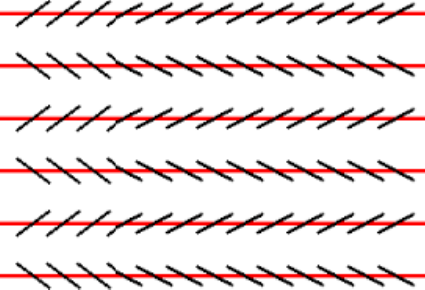}
    \caption{A version of Zöllner illusion: Those red lines are parallel, and since we know that we made a wrong intuition, all those lines still seem to be sloping and even shaking}
    \label{fig:misunderstanding}
\end{figure}

To explain this phenomenon, we designed the weight $a_s$ and $a_k$. We thought that in both human's and CTM's daily life, those kinds of visual illusion rarely happens. So since sensation and knowledgement are always highly paralleled, there is no need to pay such high attention to both dynamic sensations and static knowledgement. The reason why we call it 'static' knowledgement is that the information is stored in LTM instead of just coming from the sensor's input, to search and call that information would take time and energy, which is a waste of finite computation. \\
Aiming to avoid wasting computation, $a_k$, the weight of \textbf{$K$} should be a near-zero number unless the modeling function received some high-level instruction or unexpected stimulation.\\

\subsection{IG \&\& TG: Generating self-conscious gist and weight them}

Regarding IG and TG, we are primarily concerned with their ability to generate instructions and thoughts. In CTM, all information exists in the form of gists (in Brainish) and is transmitted based on their corresponding weights. Therefore, we propose the Gist Function and Value Function, which should be overloaded by almost all processors. Here, we focus more on the overloading of these two functions by IG and TG.\\

\subsubsection{Gist function \textbf{G}}
\ 
\newline
The MIT can generate instructions and thoughts, which means it should generate appropriate gists according to the CTM's state. So we define the gist function \textbf{G} as: 
\[[Gists] = G(S, I, K))\]
\[S=M(Sensation, Knowledge)\]
\begin{itemize}
    \item \textbf{S} stands for the state(the output of modeling function \textbf{M}), including both inner world and outer world. An elaborate description of the world is vital for generating possible gists.
    \item \textbf{K} stands for knowledgement, which refers to memories and common sense like laws of physics, laws, social regulation, and so on.
    \item \textbf{I} stands for the information that IG\&TG got from each other's previous work.
\end{itemize}

This function \textbf{G} takes the state as input, based on its knowledgement of the world, it will output a matrix of gist, which might be possible actions in IG and thoughts, answers, and questions in TG. Noticed that all those are encoded in Branish and none of them has been assigned weight yet. \\

Both IG and TG have close communication with MotW, similarly, there is also very close communication between IG and TG. When IG generates instructions, it will refer to the thoughts generated in TG, and when TG generates thinking instructions, it will also refer to the action instructions generated in IG. We assume that the Gist functions of these two processors will simultaneously(at the same tick) calculate their results, and they should both take each other's calculation results from the previous ticks as input. We can easily find examples in the human brain that conform to the above description: thoughts can affect current behavior, and past behavior can also affect current thoughts.\\

\subsubsection{Value function \textbf{V}}
\ 
\newline
All processors should have their value function, which assigns weights to chunks to measure their importance.
To be a self-conscious model, the CTM should have its values. For example, when it feels pain, it should keep away from what hurt it. 
That's why we defined the value function \textbf{V} as:
\[[weight] = V(G, S, K t) \]

\begin{itemize}
    \item \textbf{S} stands for the state of CTM.
    \item \textbf{G} stands for the matrix of gist generated by the Gist function.
    \item \textbf{K} stands for knowledgement.
    \item \textbf{t} stands for a vector of time and indicates the reception time of those gist in matrix G.
\end{itemize}
The value function \textbf{V} takes a gist matrix and the state of CTM as input, using a pre-trained classifier C, it would assign all the possible gists with their value. We believe that the value function is of great importance. Just like human beings, the distinction between value function in IG and TG could be the origin of personality and self-consciousness. \\
We believe that in the CTM, although all processors have their value functions, only a subset of them are pre-trained. "Pre-training" refers to the ability to output relatively reasonable results based on an individual's situation without any or very limited experience. That is to say, when CTM is an 'infant', it still has to make correct decisions in some simple states without former knowledgement or feedback from the environment. \\
When we say \textbf{pre-trained value function}, we refer to a pre-trained classifier  \textbf{C} and a pre-trained model \textbf{Cur}, which can be describe as :
\[C(gist) = P,\quad P \in \{0,1\}\]
\[Cur(gist) = Curiosity, \quad Curiosity \in [0, upper\_bound]\]
The classifier \textbf{C} can take gist as input and output an integer(0 or 1), to indicate whether the CTM should feel pleasure or pain to this gist. What's more, the pre-trained function $Cur(gist))$ aims at figuring out whether the CTM is curious about the gist or not and how much it is curious about it. While the CTM is too curious about some gist, it will get a higher weight from the value function. \\
While we've discussed three different kinds of attitudes that the CTM could have towards the gist, we are going to explore how to measure the level of pleasure or pain. At a CTM's birth, we prefer to default $W$, the level of pleasure and pain, to be 1. But during the CTM's lifetime, the $W$ would change according to both the CTM's state, its knowledgement, and the information from other processors(especially IG/TG) and STM.\\
The function could be described as:
\[W=W(S, K, I)\]
Based on the discussions, here we present a complete version of the value function. \\
\[ V(S, G, K, t) = W*(-1)^{C(gist)} + Cur - t\]
Noticed that a positive value refers to pleasure and a negative value refers to pain. The minus t is to make sure the earlier received gist gets a higher weight so that it won't stuck in the MotW for a long while. And the fact is the formation of a value function is a long-term procedure last the CTM's lifetime and largely depends on CTM's experience. \\

In addition, we propose that C, Cur, and W should be dynamically changing throughout the lifetime of CTM as it interacts with the environment, including both 'natural' and 'social' environments. The interaction between IG and TG may also be implicit in the changes of C, Cur, and W, which could have an impact on their Value Functions.\\

After assigning all the gist with its weight, the value function would choose which gist is more valuable that should be packed in chunks and sent to the STM. And as far as we are concerned, a gist with higher weight should be packed in chunks with a higher probability.\\

\subsection{A limited Cache \textbf{C} and Long Term Storage \textbf{L}}
\
\newline
All the LTM processors have limited memory. So it makes sense to discuss which kinds of knowledgement should be stored in their short-term memory, what we called the limited cache \textbf{C}, and which should be stored in their long-term memory \textbf{L}.\\
We decide to take MotW as an example. First, we should make a list of MotW's functions and find which of them could be done on its own. As we've discussed before, sensations and knowledgement are the two main arguments taken by MotW's modeling function. We'll talk about these two important arguments respectively. \\
\begin{itemize}
    \item [1)] About the sensations, we believe they were a group of Branish sequences continuously sent to the MoTW mainly via links for the reason that the MoTW would handle tons of data from all the sensors to model the world, the connection between  MoTW and the sensor processors should be strong enough to activate links between them. However, storing all the data could cost a huge waste of limited storage. So we think those sensations would just be stored in the cache \textbf{C}. After processing, most of them will be abandoned and only a fraction of them would leave in the long-term memory of CTM. 
    \item [2)] When it comes to knowledgement, first we clarify that it includes the memory of life and knowledge of the world. About the memory part, we prefer to believe that they aren't stored in the \textbf{L} of MoTW. In our opinion, the dream is a good experiment for this question. Although it's hard to research it, we're all convinced that people wouldn't randomly make dreams most time. Most of our dream consist of daily life, the people we know, and the places we've been to. However, unless we are really into something, we seldom make the same dream for days. And people tend to make dreams about their recent life. In our explanation, those phenomena are caused by some strong links between MoTW and other processors including IG, TG, and all other processors. Some active processor interacts with MoTW frequently during the daytime so the link remains and keep sending chunks to the MoTW. Based on those chunks, the MoTW would model a virtual world, which is our dream. 
    \item [3)] About knowledge of the world, consists of laws in different fields, for example, the physical laws. It's also a huge amount of data and can hardly be stored one hundred percent in the MoTW's long-term memory. While we're dreaming, of course, there are nearly no sensation inputs, the dream world was formed by our knowledgement. In our dreams, sometimes we could fly, armed with an Iron Man suit to fight against Thanos or even give spells like ‘Wingardium Leviosa’, so it's clear that there are no strict laws while we're dreaming. That means those laws and rules are not mainly stored in the MoTW itself(otherwise our dream will strictly obey them). Its mechanism might be similar to the memory part mentioned above for the reason that people are more likely to dream of Transformers if they've just seen the movie in the daytime.\\
\end{itemize}
After all the discussion, we conclude that neither sensations nor knowledge is stored in the MoTW's long-term memory. However, there is one special piece of information that would be stored in it, that is, what we called the basic 'feeling' of the three-dimensional world. Although people could make diverse dreams, they seldom dream of a two-dimensional world or a higher-dimensional world. We consider this special kind of feeling as an objective comprehension of the world we're living(or the CTM is living) in. It includes more fundamental laws of our world compared with all those laws of knowledgement and might be more abstract, even different from one to another.

\subsection{Further discussion}

In this section, we discussed some functions of MIT, but there are still some issues that remain. Not all chunks generated by MIT will ultimately become self-conscious content, as we have mentioned in the definition. For example, the understanding of the outer world generated by MoTW does not involve self-awareness, even if it undergoes broadcast, it can only form conscious awareness rather than self-conscious awareness.\\

Chunks generated by IG and TG may not necessarily form self-conscious awareness either. For example, a certain chunk may stimulate IG to directly establish a link with actuator processors or MoTW, or activate an existing link between them, resulting in the chunk generated by IG not entering the up-tree, but being accepted by actuators or MoTW through the link, forming sense or action. For humans, a prime example of this is the disease \textbf{PTSD}.

In addition, because we are discussing a simplified CTM with only one STM, it is far from enough to handle the amount of information that $10^{7}$ processors can generate per tick. This has led us to strengthen the links between the three processors and between the processors and sensors to ensure that they can obtain enough information. However, in reality, a true CTM should have a large number of STMs and may even be able to send and receive large amounts of chunks asynchronously. Based on this assurance, the importance of links might be reduced. From another perspective, perhaps there exists a certain type of \textbf{local STM} in a real CTM that functions similarly to the STM but is specific to certain groups of processors, like the MIT. What would that be like?\\

The discussion on this point is interesting, and we hope that one day we can obtain satisfactory results.

\section{Some explanation from illusions and disorders}
We’ve already finished defining our model and we want to get a further step on rationality. For human beings, the most classical perspective we get self-conscious experience is our body. One great source of explanations is illusions and disorders of body representation. The human brain constantly sends and receives a flow of multisensory information. The integration of this information is responsible not only for the way our body is represented but also for the way it is consciously experienced. Recent studies in cognitive neuroscience have shown that it is possible to modulate bodily self-conscious experience by experimentally changing these multisensory bodily signals. For example, Indian neuropsychologist Vilayanur Ramachandran designed an interesting experiment using a mirrored box to generate synesthesia and kinesthetic illusions in phantom limbs.
\subsection{Synaesthesia in phantom limbs}
\textbf{Phantom Limbs} refers to a phenomenon in that amputees experience persistent and vivid sensations in their physically absent limb, typically appearing after the accidental loss\cite{phantom_sensation_basic}.\\

Mirrors generate a symmetrical image for the object in front of it. Using this property we can generate phantom limbs. Ramachandran and colleagues constructed a “virtual reality box” which is a cardboard box with a mirror vertically inserted in the middle. They cut two holes in the enclosure of each side then the patient was told to put both his or her arms into the holes. Next, the patient was asked to observe his or her healthy hand in the mirror. Then ask the patient to do symmetrical movements on both sides and see what happened\cite{phantom_box}.\\

\begin{figure}[htp]
    \centering
    \includegraphics[width=8cm]{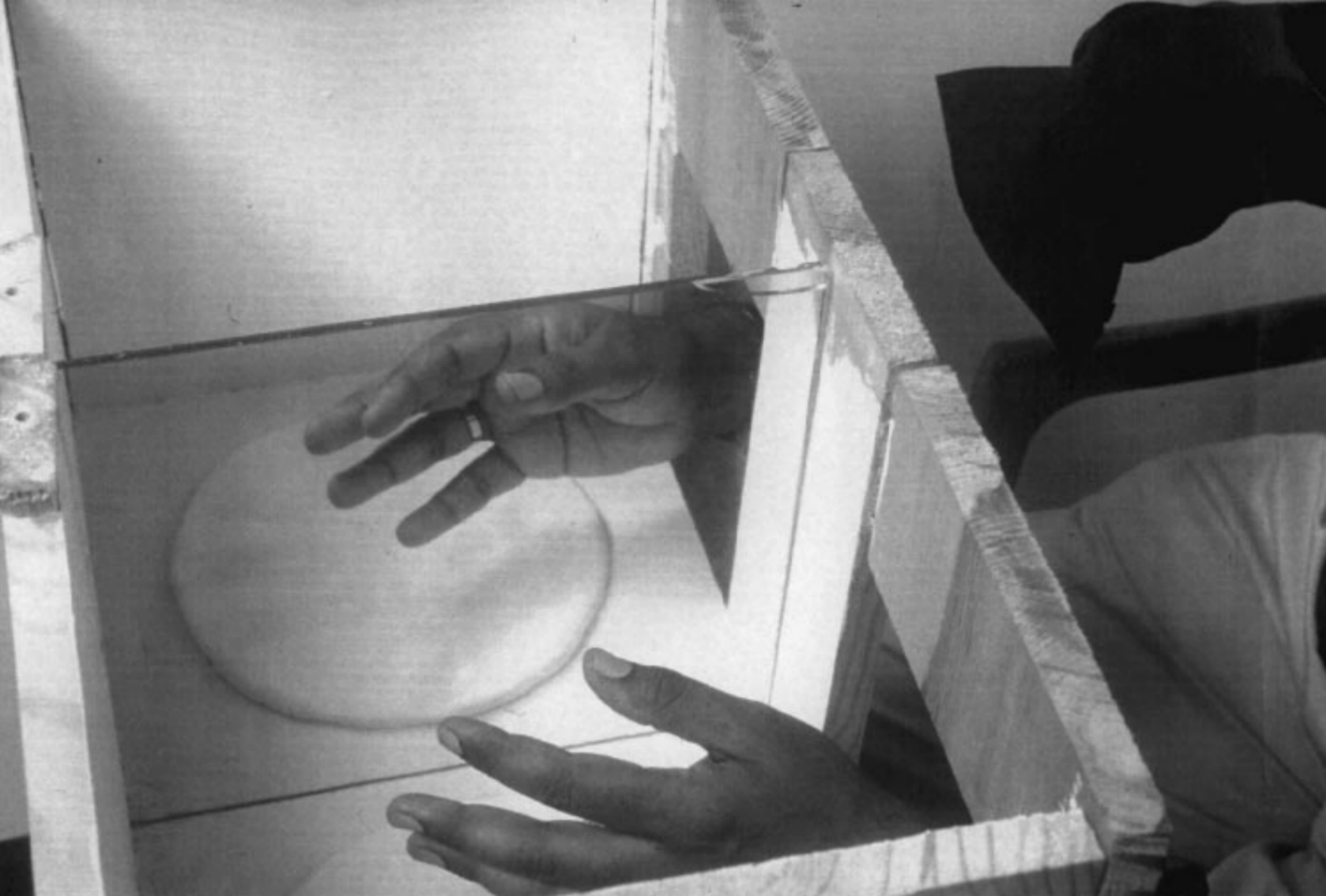}
    \caption{Virtual reality box made by Ramachandran and colleagues. When amputees patient put their hands through the holes in the enclosure, they can observe an image of the whole hand generated by their healthy hand.}
    \label{fig:misunderstanding}
\end{figure}

Among 10 patients, six patients perceived phantom when the normal hand was moved. For example, Mr. R.T. was an intelligent, 55-year-old engineer who had an infiltrating sarcoma in his left arm that produced a painful ulnar nerve palsy. Six months later his arm was amputated 6" above the elbow. When the researchers asked him to simultaneously send motor commands to both hands as if to perform mirror symmetrical movements, e.g. clenching and unclenching of the fist, extension, and flexion of the wrist, or circular movements. The very first time he tried this and the patient exclaimed with considerable surprise, that all his movements had come back that he now vividly experienced muscle and joint movements in his phantom!  And this can be even repeated. The patient R.L. tried the same experiment six times on different occasions and he perceived a phantom every time\cite{phantom_story}.\\ 

Back to CTM and our model, let’s consider what will happen if a CTM wants to clench and unclench its hand. Once the CTM has the mind to clench its hand, MotW will first respond to it. Modeling function $M$ will first generate an output of its hand which is called hand state. Then the output will be passed to processors $IG$ and $TG$ which generates possible gist based on the hand state and knowledgement $K$. Next, usually it gives a chunk and it is quickly broadcast through STM to the related processor responsible to implement the action for conscious actions. And finally, the CTM clenches its hand. So if a CTM without a hand wants to clench its hand, the whole process we mentioned above will be stuck at the instructions-generating processor $IG$. \\

How to explain phantom limb in a CTM? The first idea that comes to our mind is that the patient has had a healthy hand and they know what it is like to be to use a healthy hand. Similarly, an amputee CTM has got experience with how to use a healthy hand and its feedback. Once when it observes a mirror image of its hand, it gives the modeling function stimulation and generates a normal state of the hand. The state output is then received by $IG$. The key point here is $IG$ call the memory(or better say knowledge) on using the hand $K$ then integrated inputs and normally generate hand actions. \\

But here comes another question,  does this action-generating process based on acquired experience about using a hand? If we want to answer the question we need to find some experiments on phantom limb phenomena that happen in congenital amputees. \\

Ronald Melzack and colleagues did an experiment among a group of people with a congenital limb deficiency or amputation in early childhood\cite{early_child}. The result shows that around 20\% of people born with congenital limb deficiency develop a phantom of the missing limb and around 50\% of children who lose a limb at the age of 5 years or younger develop a phantom limb. Though the descriptions of phantoms provided by young children cannot be trusted, perhaps due to the belief that children cannot differentiate between imagination and reality, or they simply want to appear ‘normal’ or that they are more suggestible than adults. But it shows that phantom limb phenomena can happen in congenital amputees.
It shows that we are born with some basic knowledge of how to use our bodies. Compare to the 60\% proportion in VS Ramachandran’s experiment 20\% is a low level. But from a dynamic perspective, as the age at that people get injured grows up, the proportion is increasing. So we can say we are born with basic knowledge of our body and it is supplemented and completed as people grow up.\\

\begin{figure}[htp]
    \centering
    \includegraphics[width=8cm]{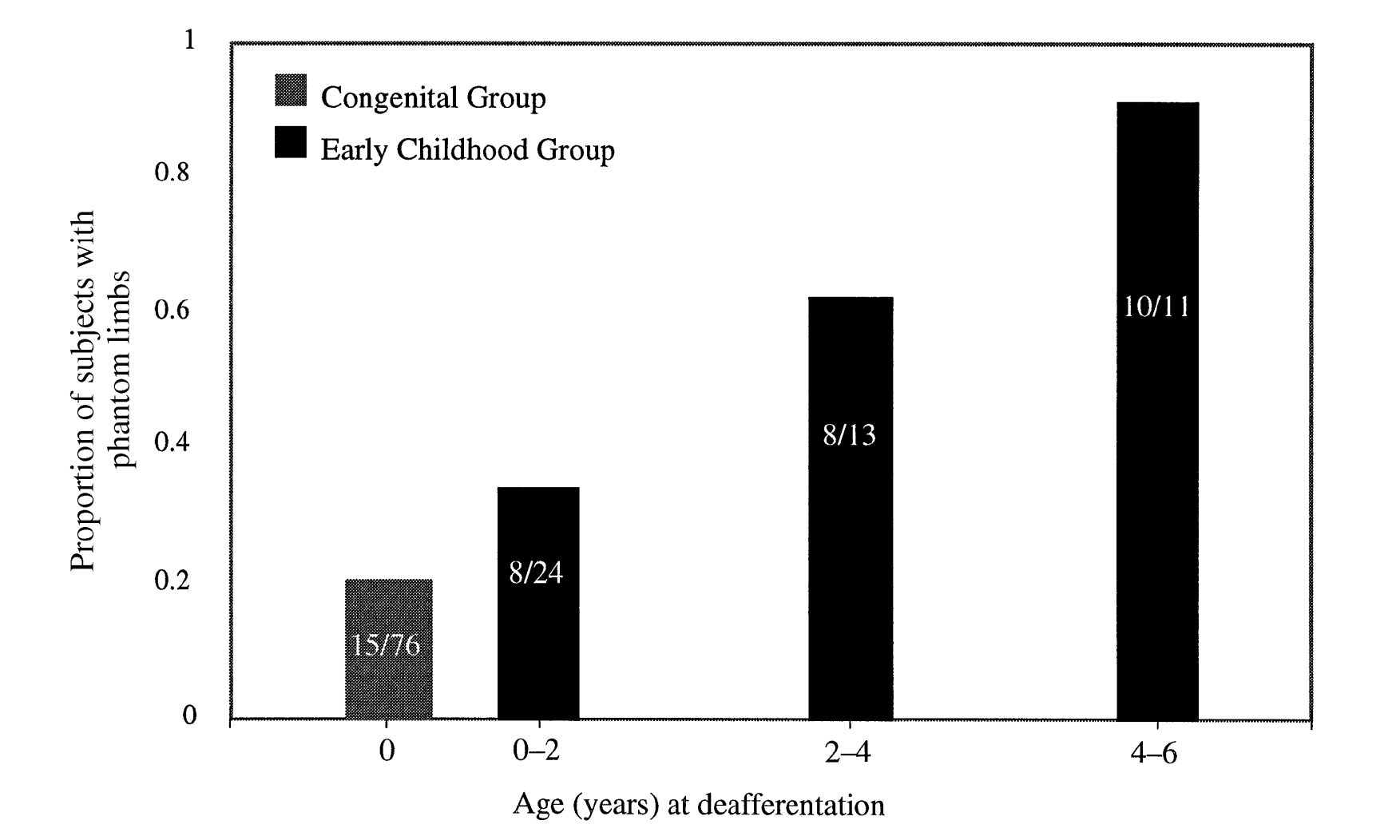}
    \caption{Relationship between occurrence of phantom limbs and age at the time of sensory deafferentation. The early amputation group was divided into three groups (deafferentation at 0–2, 2–4, and 4–6 years).}
    \label{fig:misunderstanding}
\end{figure}

Using this conclusion on our model we can say the CTM is born with some innate knowledge of its body including its usage methods, structure information, and appearance features. As time pass by after CTM spends a period of a lifetime, the knowledge becomes more complete and important by increasing weight $a_k$ in modeling function $f$.\\
In order to explore more evidence of the innate component of knowledge, let's look at another experiment designed by Brugger et al\cite{phantom_AZ}. They introduced a vividness rating on a 7-point scale that showed highly consistent judgments across sessions for their subjective experience of phantom limbs.
In the experiment, the subject was a 44-year-old university-educated woman who was born without forearms and legs. In her memory, she has experienced mental images of forearms and legs, but these were not all in the same subjective reality in her experience. The exciting result of functional magnetic resonance imaging of phantom hand movements showed no activation of the primary sensorimotor areas, but of the premotor and parietal cortex bilaterally. Transcranial magnetic stimulation of the sensorimotor cortex consistently elicited phantom sensations in the contralateral fingers and hand. In addition, premotor and parietal stimulation evoked similar phantom sensations, albeit in the absence of motor-evoked potentials in the stump. \\
\begin{figure}[htp]
    \centering
    \includegraphics[width=5cm]{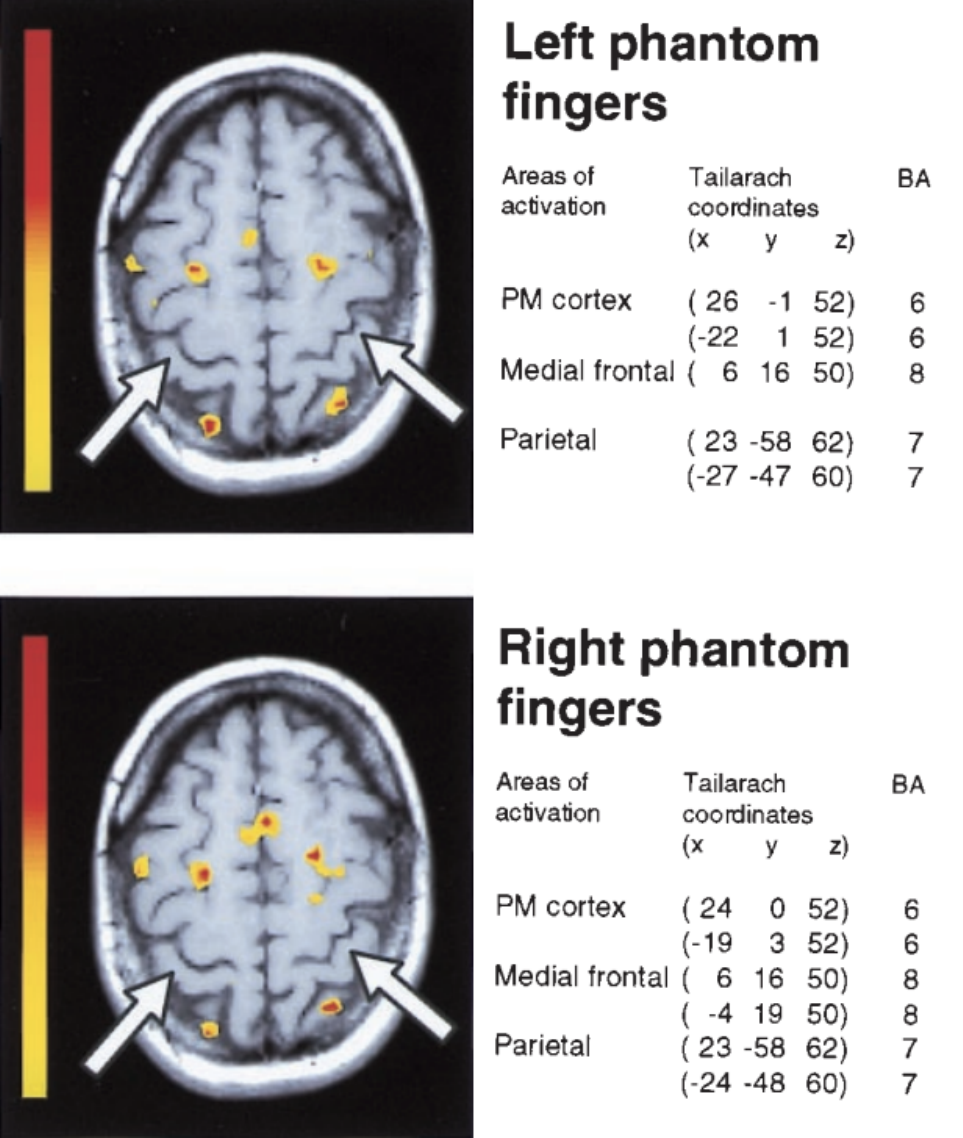}
    \caption{Cortical activation areas during self-paced movements of phantom fingers. Two representative sections through the sensorimotor cortical hand areas during fingers-to-thumb opposition with the left (Top) and right (Bottom) phantom fingers. }
    \label{fig:misunderstanding}
\end{figure}
These data show strong evidence that body parts that were never physically developed can be simulated in sensory and motor cortical areas. Similar with human brain, in CTM we simulate this process by sensation processors, MotW and IG processor. The data strongly support our mechanism and show a new method for further developing the model.

\section{Conclusion}
In this paper, we prove that the definition of CTM's conscious is convincing. Based on the powerful model we can develop mechanisms and better understand human conscious processes. What's more, we present a possible group of processors dealing with self-consciousness, we believe that this very special group of processors plays a key role in generating self-consciousness. Human consciousness has always been a source of ideas and evidence\cite{illusion_book}. We try to make our model simple but as we consider more details and examples of human consciousness, we build more processors that have specific functions. Finally, we use phantom limbs to explain our model and get a better understanding of the workflow. Surprisingly, the result of functional magnetic resonance imaging of phantom hand movements also shows supporting and intuitive evidence for the structure of the model. Because of time limitations, we didn't find more illusions and disorders, however, we point out an idea of self-conscious mechanism in CTM and methods to prove the correctness. Hoping that will be helpful for further research.

\bibliography{cite}

\end{document}